\documentclass[prl,aps,twocolumn,superscriptaddress,showpacs]{revtex4}
\usepackage{graphicx}
\def\(({\left(}
\def\)){\right)}
\def\[[{\left[}
\def\]]{\right]}
\begin{document}

\title{Temperature chaos, rejuvenation and memory in Migdal-Kadanoff
spin glasses}
\author{M. Sasaki}
\affiliation{
Institute for Solid State Physics, University of Tokyo, 
Kashiwa-no-ha 5-1-5, Kashiwa, 277-8581, Japan}

\author{O.C. Martin}
\affiliation{
Laboratoire de Physique Th\'eorique et Mod\`eles Statistiques,
b\^at. 100, Universit\'e Paris-Sud, F--91405 Orsay, France.}

\date{\today}

\begin{abstract}
We use simulations within the Migdal-Kadanoff approach to probe 
the scales relevant for rejuvenation and memory 
in Ising spin glasses. First we investigate scaling laws 
for domain wall free energies and extract the chaos overlap 
length $\ell(T,T')$. Then we perform out of equilibrium simulations 
that follow experimental protocols. We find that: 
(1) a rejuvenation signal arises at a length scale significantly 
smaller than $\ell(T,T')$; (2) memory survives even if equilibration 
goes out to length scales larger than $\ell(T,T')$. 
Theoretical justifications of these phenomena are then considered.
\end{abstract}

\maketitle

Two of the most spectacular experimental properties of spin 
glasses are ``rejuvenation'' and ``memory'' (see~\cite{DupuisVincent01}
and references therein). Both are out of equilibrium phenomena
that arise in slowly relaxing systems~\cite{Struik78}, so 
understanding such properties is of great importance. Qualitatively,
when a spin glass approaches equilibrium, it ``ages'', reducing
its susceptibility, that is its response to external perturbations.
However, if one lowers the temperature after aging, one sees
a restart or rejuvenation of the susceptibility, while
memory of the previous aging can be retrieved 
upon heating back! From
a theoretical point of view, rejuvenation must appear if
there is ``temperature chaos'', that is if the 
spin polarizations at two temperatures
$T$ and $T'$ are decorrelated~\cite{BrayMoore87} beyond a characteristic
length scale $\ell(T,T')$. 
Moreover, temperature chaos is compatible with 
memory through the presence of ghost domains~\cite{YoshinoLemaitre00}.
However, temperature chaos is {\it not} clearly seen 
in Monte Carlo simulations~\cite{BilloireMarinari00,BilloireMarinari02}
and estimates of the chaos length $\ell(T,T')$ give very large
values~\cite{AspelmeierBray02}, seemingly much larger
than the length scale $\ell_R$ on which rejuvenation appears experimentally.

Our purpose here is to find the scales relevant
for rejuvenation and memory
in Ising spin glasses. We use
Migdal-Kadanoff (MK) lattices whose
exact renormalization~\cite{SouthernYoung77}
allows one to measure equilibrium quantities on
large {\it length} scales.
Furthermore, it also allows for efficient dynamical simulations
at very long {\it time} scales, enabling us to extract
the length $\ell_R$ which is relevant for rejuvenation.
The outline of the paper is as follows. 
First, we define the MK lattices. 
Second we investigate {\it equilibrium} chaos: after extracting
$\ell(T,T')$, we determine how the distribution
of two-temperature overlaps changes with lattice size. 
Third, we show how (renormalized) dynamics can be used to probe
rejuvenation and memory on very long length and time scales. Finally, we
perform out of equilibrium
measurements that follow standard experimental protocols;
these signal rejuvenation 
even if temperature chaos is very weak
and show that memory is preserved even if the equilibrated length 
scale is much larger than 
$\ell(T,T')$. The current theoretical frameworks
(see~\cite{BerthierBouchaud02,JonssonYoshino02} 
and~\cite{YoshinoLemaitre00}) partially
account for these properties as explained in the discussion section.

\begin{figure}[bb]
\includegraphics[angle=0,width=\columnwidth]{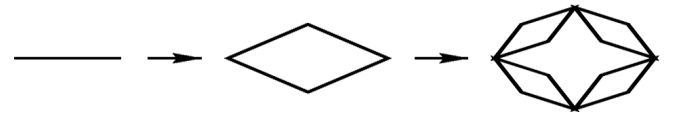} 
\caption{Construction of a hierarchical MK lattice ($b=2$).}
\label{fig:mk}
\end{figure}

\paragraph*{The model ---}
We consider MK lattices following 
the standard real space renormalization group
approximation~\cite{SouthernYoung77} to the
Edwards Anderson (EA) model~\cite{EdwardsAnderson75}.
The recursive construction of such hierarchical lattices is
described in Fig.~\ref{fig:mk}; edges are
replaced by $2 b$ edges so the ``length'' of the lattice
is multiplied by $2$. We call generation ``level'' the order
of the recursion and $G$ 
the total number of these. Then the lattice length $L$ is
$2^G$ and the number of bonds is $(2 b)^G$ (which is also
roughly the number of sites); one can thus identify
$1 + \ln b / \ln 2$ with 
the dimension of space on such a lattice.
When all the edges are constructed, each is assigned
a random coupling $J_{ij}$. Similarly,
on each site $i$ we put an Ising spin $S_i = \pm 1$. 
The Hamiltonian is
\begin{equation}
\label{eq:H}
H_J(\{S_i\}) = - \sum_{<ij>} J_{ij} S_i S_j
\end{equation}
where the sum is over all the nearest 
neighbor spins of the lattice. The MK approach leads to
accurate values for the spin glass stiffness exponent
$\theta$ and for
the lower critical dimension; furthermore it
exhibits temperature 
chaos~\cite{BanavarBray87,NifleHilhorst92,AspelmeierBray02}. We thus 
feel it is a good starting point for studying the mechanisms
of rejuvenation and memory in spin glasses.
All of the work presented here will be for 
three dimensions ($b=4$) with couplings $J_{ij}$ taken
from a Gaussian of mean $0$ and width $1$.
The model then undergoes a spin glass transition
at $T_c \approx  0.896$~\cite{NifleHilhorst92}.

\paragraph*{Chaos in domain wall free energies ---}
Bray and Moore~\cite{BrayMoore87} were the first to study
the temperature dependence of {\it domain wall} free energies.
Call $F_J^{{\rm DW}}$ the free energy of a domain wall 
for a given disorder obtained by forcing the outer-most spins
of the MK lattice to be anti-parallel instead 
of parallel. ($F_J^{{\rm DW}}/2$ is 
the effective coupling between
these spins.) When $L$ grows, $F_J^{{\rm DW}}(T)$ and 
$F_J^{{\rm DW}}(T')$ become 
decorrelated if $T \ne T'$, so the
linear correlation coefficient~\cite{NifleHilhorst92}
\begin{equation}
C^{{\rm DW}}(L,T,T') = \frac{\overline{F_J^{{\rm DW}}(T) F_J^{{\rm DW}}(T')}}
{\sigma(T) ~~ \sigma(T')}
\label{eq:correlation_coeff}
\end{equation}
goes to zero at large $L$.
(In this definition, 
${\overline{ \cdots }}$ is the disorder average,
$\sigma$ is the standard deviation of $F_J^{{\rm DW}}$,
and we have used the fact that $\overline{F_J^{{\rm DW}}}=0$.)
\begin{figure}[bb]
\includegraphics[width=\columnwidth]{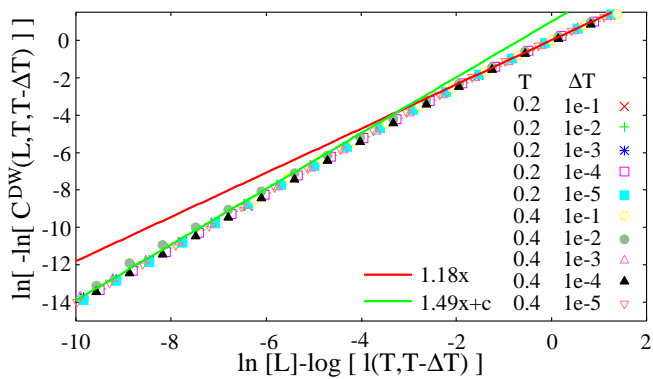} 
\caption{Plot of 
$\ln \{ -\ln [ C^{{\rm DW}}(L,T,T-\Delta T) ] \}$
to show the data collapse. 
The lines are the weak and strong chaos limits.}
\label{fig:CLscaling}
\end{figure}
We are not aware of any study describing
{\it how} $C^{{\rm DW}}$ vanishes when $L$ grows. We thus 
computed $F_J^{{\rm DW}}$ for
a large number of MK lattices and then
estimated $C^{{\rm DW}}$. Defining $\ell(T,T')$ as the value of $L$ where
$C^{{\rm DW}}=1/e$, our
data fall on a single curve when using
the scaling variable $L/\ell(T,T')$:
\begin{equation}
C^{{\rm DW}}(L,T,T') \sim f[L/\ell(T,T')]
\label{eq:CLscaling}
\end{equation}
This is illustrated in Fig.~\ref{fig:CLscaling}.
As expected, $\ell(T,T')$ goes as
$|T-T'|^{-1/\zeta}$ where
$\zeta$ is the chaos exponent~\cite{BanavarBray87}; for our
system, $\zeta = d_{\rm s}/2 - \theta \approx 0.745$.
As shown in ref.~\cite{NifleHilhorst92}, 
in the weak chaos limit $L \ll \ell(T,T')$, 
$1 - C^{{\rm DW}} \approx \left[ L/\ell(T,T') \right]^{2 \zeta}$.
In the strong chaos limit, 
$L \gg \ell(T,T')$, we find that
the scaling function $f(x)$ behaves as
$\exp(-x^\alpha)$, with $\alpha =1.18 \pm 0.02$.

\paragraph*{The two temperature $P(q)$ ---} Beyond the
length scale $\ell(T,T')$, 
domain wall free energies will often have different signs
at $T$ and $T'$. As a consequence, the spin orderings
will be different as can be made quantitative
by considering two-temperature ``overlaps''.
Let $q$ be the overlap of two configurations ${\cal C}$ 
and ${\cal C}'$ taken in equilibrium (at $T$ for $\cal C$ and
at $T'$ for ${\cal C}'$). Temperature chaos implies that 
$P_{TT'}(q)$, the disorder averaged distribution of such overlaps, 
tends towards a delta function in $0$. However, 
this behavior has not been seen in the 
Sherrington-Kirkpatrick (SK) model nor
in EA spin glasses~\cite{BilloireMarinari00,BilloireMarinari02}.
It is therefore interesting to see how 
$P_{TT'}(q)$ behaves in MK spin glasses where 
temperature chaos arises for sure and $\ell(T,T')$ is known. 

\begin{figure}[bb]
\includegraphics[width=\columnwidth]{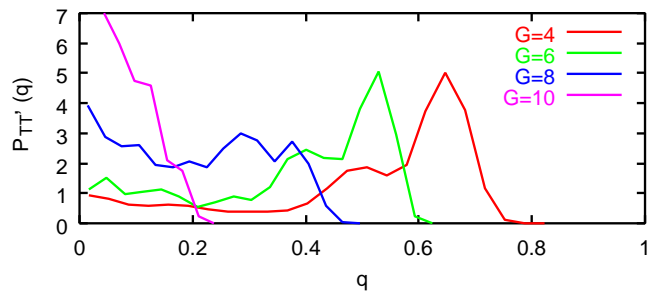} 
\caption{Two-temperature overlap probability distribution
$P_{TT'}(q)$ when $T=0.7$, $T'=0.4$; $L=2^G$.}
\label{fig:pofq}
\end{figure}

In Fig.~\ref{fig:pofq} we show $P_{TT'}(q)$ determined numerically 
in the case $T=0.7$, $T'=0.4$ for which $\ell(T,T')=2^{7.8}$.
When $L=2^G\le 2^6$, $P_{TT'}(q)$ for $T\ne T'$ and $T=T'$ are similar:
there is a main peak at a ``large'' value of $q$ and a broad
tail towards lower $q$; furthermore
the peak position shifts to lower $q$ with increasing $G$.
The differences from the case $T=T'$ are that 
the distributions have a shoulder and also
a clear local maximum at $q=0$. This behavior is close to 
what has been observed in the SK model~\cite{BilloireMarinari02}.
Now for larger $G$'s, the shoulder takes over, 
and starting with $G=9$, $P_{TT'}(q)$ has a single peak,
located at $q=0$. Note that $G=6$ corresponds to
very large $L$ and so the asymptotic behavior is not likely to be seen soon
in the EA model.

\paragraph*{Exploiting renormalization for dynamical quantities ---}
Temperature chaos has often been used to explain rejuvenation,
but temperature chaos is not necessary for rejuvenation. 
In particular, rejuvenation arises in generalized 
random energy models~\cite{SasakiNemoto00}. 
However, a study~\cite{JonssonYoshino02} has suggested 
that the influence of temperature chaos may be visible 
even if the domain size (or equilibrated length scale)
reached experimentally is much smaller 
than the overlap length $\ell(T,T')$.
Thus we ask here what is the 
relation between the (equilibrium) overlap length 
and the length scale $l_{R}$ which is relevant for rejuvenation. 
In the framework of Migdal-Kadanoff lattices,
we can address this question because one may go to
long time scales as follows. 

Suppose we focus on a time window
$t_{\rm min}\le t \le t_{\rm max}$. Between $t=0$ and
$t=t_{\rm min}$ the system has had time to equilibrate up to
the length scale $l(t_{\rm min})$; essentially all out of equilibrium
physics comes 
from larger length scales. On the MK lattice, this means
that the spins whose generation ``level'' is larger than $G_{\rm min}$
(with $2^{G-G_{\rm min}} = l(t_{\rm min})$) are in local 
equilibrium; the other spins have dynamics that is
well described by the effective Hamiltonian at the
generation $G_{\rm min}$. In practice, 
we implement this idea as follows.
First we generate a large number of bare couplings 
from a Gaussian of mean $0$ and width $1$. 
Then, we use renormalization 
to produce an ensemble of effective couplings. This process
is iterated $\textrm{NRG}\equiv G-G_{\rm min}$ times. 
($\textrm{NRG}$ is for the number of renormalization group transformations.) 
The final effective couplings are then used to create 
a MK lattice of size $2^{G_{\rm min}}$. After that, 
we simply do standard Monte Carlo. 
Note that one Monte Carlo Sweep (MCS) on the renormalized
lattice corresponds to a (huge) number of sweeps
on the non-renormalized lattice, in fact to the
number needed to equilibrate on the length scale $2^{\textrm{NRG}}$. 

\begin{figure}[bb]
\includegraphics[angle=0,width=7.6cm]{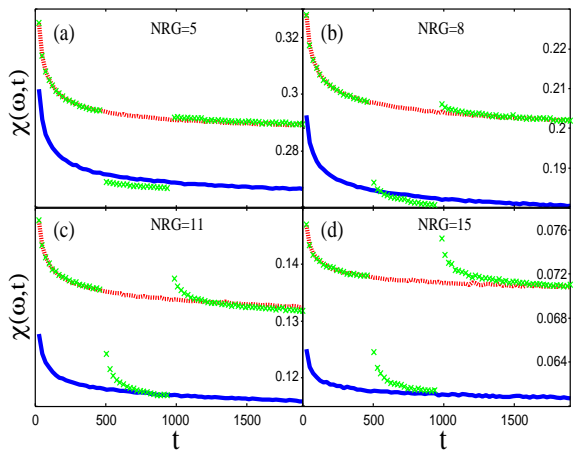}
\caption{Isothermal $\chi(\omega,t)$ at $T$ (broken line), at $T-\Delta T$ 
(solid line), and $\chi(\omega,t)$ 
with a negative T-cycle (crosses) for $\textrm{NRG}=5$, 8, 11 and 15. 
The average is from $10^4$ samples.}
\label{fig:Tcycle}
\end{figure}

\paragraph*{Memory and Rejuvenation ---}
We use the standard temperature cycling protocol and measure
a quantity similar to the ac-susceptibility defined as~\cite{KomoriYoshino00}
\begin{equation}
\chi(\omega,t)=\frac{1-Q(t+\frac{2\pi}{\omega},t)}{T}
\end{equation}
where $Q(t,t')\equiv \sum_i \langle S_i(t) S_i(t') \rangle / N$.
(Note that $Q$ is a dynamical generalization of the $q$ previously
discussed.)
The period ${2\pi}/{\omega}$ of our ac-field is 16 MCS. 
Every MCS updates all spins once.
An alternative choice is to sweep the bonds, updating their end spins
as in~\cite{RicciRitort00}; we 
have checked that the results are hardly affected by the method
used. The simulations were done on Migdal-Kadanoff lattices with 
four generations using $0\le \textrm{NRG} \le 15$. 
In Fig.~\ref{fig:Tcycle} we
show the isothermal $\chi$ at $T$, at $T-\Delta T$, and also 
$\chi$ for a $T \to T-\Delta T \to T$ temperature cycle.
We used $T=0.7$ and 
$\Delta T=0.05$. Since we calculate renormalized couplings 
at $T$ and $T-\Delta T$ from the {\it same} set of bare couplings, 
they are highly correlated when $\textrm{NRG}$ is small. However, 
their correlation vanishes for large $\textrm{NRG}$ due to temperature chaos. 
The direction of each spin at $t=0$ is chosen randomly 
with equal probability, corresponding to
a quench from an infinitely high temperature at an infinite rate. 
We hereafter denote $\chi$ with a T-cycle as $\chi_{\rm cycle}$ 
and the isothermal $\chi$ at $T$ as $\chi_{\rm iso}(T)$. 

For small $\textrm{NRG}$, $\chi_{\rm cycle}$ is 
below $\chi_{\rm iso}(T-\Delta T)$ 
in the second stage, as illustrated in Fig.~\ref{fig:Tcycle}(a). 
This means that the former is {\it older}
than the latter and that equilibration is accelerated 
with increasing temperature; there is no rejuvenation here.
On the other hand, $\chi_{\rm cycle}$ and $\chi_{\rm iso}(T)$
almost overlap in the third stage. 
This means that the effective aging time $t_{\rm eff}$, 
that is the equivalent time the system would have had to spend
at $T$ rather than at $T-\Delta T$ to return to the same curve,
is close to the actual window time $t_{\rm w2}$ spent at $T-\Delta T$. 
It is often observed in experiment that $t_{\rm eff}$ becomes 0 
for large $\Delta T$, corresponding to {\it perfect memory}. 
In our simulations, $\Delta T$ is relatively small and so we get
partial memory
for all $\textrm{NRG}$. This general trend of no rejuvenation yet memory 
with $t_{\rm eff}\approx t_{\rm w2}$ arises until $\textrm{NRG}\approx 7$. 

Fig.~\ref{fig:Tcycle}(b) shows that 
a sign of rejuvenation appears around $\textrm{NRG}\approx 8$, 
i.e., $\chi_{\rm cycle}$ is above $\chi_{\rm iso}(T-\Delta T)$ 
at the {\em beginning} of the second stage.  
However, at later times of this second stage, the
cycling curve goes {\em below} the $\chi_{\rm iso}(T-\Delta T)$ curve.
The crossing of these curves has been observed 
experimentally~\cite{Dupuis02},
and will be examined in the discussion section.
Finally at the beginning of the third stage, the cycling data has
significant deviations from the isothermal data. The conclusion for
this figure is that signs of
rejuvenation emerge on smaller length scales than expected 
because the linear correlation coefficient is
still large when $\textrm{NRG}=8$, $C^{{\rm DW}} = 0.965$. 
($\ell(0.7,0.65) \approx 2^{12}$ in this case.)
This result is consistent with that in~\cite{PiccoRicci01b} 
where rejuvenation was observed in the $3$-dimensional 
EA spin glass model when reversing the sign of $5\%$ of the
couplings. We have also compared 
$\chi_{\rm cycle}$ in the second stage to 
$\chi_{\rm iso}(T-\Delta T)$ shifted to the right by $t_{\rm w1}$, 
the time of the first stage, and found that 
{\it perfect rejuvenation} (i.e., complete overlap of the two curves) 
arises only when $\textrm{NRG}\ge 13$ where the effective couplings 
at the two temperatures are very decorrelated 
($C^{{\rm DW}} \leq 0.10$). 

On general grounds, one may expect $t_{\rm eff}$ to be smaller 
than $t_{\rm w2}$ 
because the temperature is lower in the second stage. However, 
when $\textrm{NRG}=10$, $11$ and $12$, 
quite surprizingly, we find that the cycling data in the 
third stage are 
slightly {\it below} the isothermal data, meaning that  
$t_{\rm eff}>t_{\rm w2}$; this is illustrated in Fig.~\ref{fig:Tcycle}(c). 

Finally, Fig.~\ref{fig:Tcycle}(d) shows the 
case $\textrm{NRG}=15$. The effective 
couplings at the two temperatures are now completely decorrelated.
As a consequence, strong relaxation 
is observed not only in the second stage but also in the third stage. 
Now an interesting question is whether memory remains or not. 
We have thus compared $\chi(\omega,t)$ 
in the third stage to that in the first stage, and found
that the former is clearly {\it older} than the later. 
This is consistent with the prediction of ref.~\cite{YoshinoLemaitre00} 
that memory survives even if there is complete temperature chaos, 
i.e., $\ell(T,T')=0$.

Now for a few remarks. First, we have also performed
{\it positive} T-cycling simulations, going 
from $T=0.7$ to $T+\Delta T=0.75$ 
in the second stage. The qualitative behavior is similar,
the main difference being that
equilibration is accelerated by the cycling.
In particular, when $\textrm{NRG}$ is not too large,
$\chi(\omega,t)$ in the third stage and at large times
is below the isothermal data. 
This kind of behavior has been observed in
glassy systems like polymer glasses~\cite{BellonCiliberto99}, 
but not yet in spin glasses. 
Second, we have used the same $\textrm{NRG}$
at $T$ and $T-\Delta T$ 
but really $\textrm{NRG}$ should decrease with temperature  
because the ordering is then slower.
A temperature dependence of $\textrm{NRG}$ will reduce $t_{\rm eff}$ 
which is unexpectedly large in our simulations. 
Moreover, it will cause separation of
length scales, an important ingredient for memory and
rejuvenation~\cite{BouchaudDupuis01}. 

\paragraph*{Discussion and conclusions ---}
One of our main findings in this work is that 
a signal of rejuvenation arises on scales $\ell_R$ much smaller than 
the equilibrium overlap length $\ell(T,T')$.
(In Fig.~\ref{fig:Tcycle}, 
rejuvenation transpires even though $\ell(T,T') / \ell_R \approx 2^4$.) 
How can one interpret this result? Within
the droplet picture, 
a small fraction of droplets of size $l$ are fragile~\cite{BrayMoore87}
against temperature variation even when $l \ll \ell(T,T')$. 
In our case, we equilibrate out to scales
$l_{\rm eq}$; beyond $l_{\rm eq}$, one has domain walls, the positions
of which are time dependent and sensitive to temperature changes.
The strong relaxation of $\chi_{\rm cycle}$
we see in the second stage can thus be interpreted as the re-ordering 
of the spins either inside fragile droplets or on the boundaries of
the (out of equilibrium) domain walls. However, it is also possible
to interpret the rejuvenation we see without appealing
to temperature chaos. Indeed, in the picture of~\cite{BouchaudDupuis01},
there is no chaos; nevertheless, temperature cycling
modifies the Boltzmann weights, leading
to rejuvenation on short length scales.

A second surprizing feature we found was the crossing
of $\chi_{\rm cycle}$ and $\chi_{\rm iso}(T-\Delta T)$.
(See Fig.~\ref{fig:Tcycle}(b).) Interestingly, such a
crossing behavior has been seen in experiments~\cite{Dupuis02}.
Do the different theoretical frameworks predict such a crossing? 
In the droplet picture, rejuvenation in Fig.~\ref{fig:Tcycle}(b) 
is attributed to fragile droplets. However, 
the equilibrium state is still robust in most regions
because the occurrence of such droplets is rare 
at this length scale, $l=2^{8}$. 
Moreover, if we compare the equilibrated length scale 
at $t_{\rm w1}$ in the T-cycling case with that in the 
isothermal case of $T-\Delta T$, the former is larger 
than the latter because equilibration is accelerated with 
increasing temperature. Therefore, $\chi_{\rm cycle}$ eventually 
goes below $\chi_{\rm iso}(T-\Delta T)$ after the re-ordering of 
fragile droplets progresses sufficiently.
The picture of~\cite{BouchaudDupuis01} also gives natural interpretation 
of the crossing. First, the positions of
pinned domain walls are determined hierarchically:
the structure at large scales is associated with large energies,
that at small scales is associated with small energies.
Second, one reaches the correct large scale structure faster by
aging at $T$ rather than at $T-\Delta T$ since
barriers are overcome more easily. (It is important
that the large scale structure be the same at the two temperatures.)
As a result, after transient reconstructions 
on smaller length scales (rejuvenation), 
$\chi_{\rm cycle}$ will cross $\chi_{\rm iso}(T-\Delta T)$.

Last, our system exhibits memory which 
persists even when the equilibrated scale is much {\it larger}
than $\ell(T,T')$. This result is completely compatible with 
the prediction by Yoshino {\it et al.}~\cite{YoshinoLemaitre00}. 
They showed that even if $\ell(T,T')\approx 0$, 
the equilibration in the second stage just injects uncorrelated 
short-range noise into the long-range ordering developed 
during the first stage; 
memory is then retrieved in the third stage 
after transients associated with removing this noise. 

We thank J.-P. Bouchaud, A. Pagnani, and E. Vincent for discussions, 
as well as H. Yoshino and his collaborators who have been working 
with a similar approach~\cite{Scheffler03}. 
This work was partially supported by the IT-program of Ministry 
of Education, Culture, Sports, Science and Technology. Some of 
the computations were carried out on the Machikaneyama PC 
cluster system (URL: http://www.mhill.org/) 
in the Large-scale Computational Science Division, Cybermedia Center, 
Osaka University. M. S. was partially supported by the 
Japan Society for the Promotion of Science for Japanese Junior Scientists. 
M. S. acknowledges support from the MENRT while he was in France. The 
LPTMS is an Unit\'e de Recherche de 
l'Universit\'e Paris~XI associ\'ee au CNRS.

\bibliographystyle{prsty}
\bibliography{references}

\end{document}